\documentclass[aps,prb,twocolumn, reprint]{revtex4-2}
\usepackage[pdftex, colorlinks=true, linkcolor=blue, citecolor=blue, urlcolor=blue]{hyperref}
\usepackage{graphicx}
\usepackage{aurical}
\usepackage{amsmath, amssymb}
\usepackage[T1]{fontenc}
\usepackage[version=3]{mhchem}
\usepackage{braket}
\graphicspath{{./figures/}}
\DeclareGraphicsExtensions{.png,.pdf}

\newcommand{\una}{Universit\'{e} de Nouakchott, Facult\'{e} des Sciences et Techniques, D\'{e}partement de Physique, Avenue du Roi Fai\c{c}al, 2373, Nouakchott, Mauritania}

\begin{document}

\title{A Purely Magnetic Route to High-Harmonic Spin Pumping}

\author{O. Ly}
\email{ousmanebouneoumar@gmail.com}
\affiliation{\una}

\begin{abstract}
Spin pumping provides a fundamental route for dynamical spin transport, yet in its conventional form it produces only linear spin responses at the driving frequency. Recent studies have shown that spin–orbit coupling (SOC) can lift these restrictions and enable strongly nonlinear spin and charge currents. Here, we propose a distinct mechanism for high-harmonic spin pumping that operates independently of spin–orbit interactions. We demonstrate that purely magnetic structures can sustain high-harmonic spin currents when an additional static magnetic order parameter is introduced perpendicular to the cone axis of a precessing magnetization. This secondary magnetic order qualitatively reshapes the dynamical response, giving rise to a cascade of higher harmonics in the pumped spin current in the absence of SOC. Our results establish a SOC-independent route to ultrafast and nonlinear spin pumping rooted solely in magnetic structure and dynamics.
\end{abstract}

\maketitle

\section{Introduction}

Spin pumping has been a cornerstone of spintronics since its formulation in the early 2000s \cite{Tserkovnyak2002b,Tserkovnyak2002}. It provides a mechanism by which a time-dependent magnetization injects a spin current into an adjacent conductor—most commonly a normal metal—leading to enhanced Gilbert damping and enabling electrical detection via spin–charge conversion mechanisms where spin currents can be detected as transverse charge voltages \cite{Saitoh2006, Ando2011PRL, Mosendz2010PRL, Wei2014}.
The effect has been explored across a broad range of material platforms and device geometries, including ferromagnet/normal-metal bilayers and low-dimensional systems \cite{Saitoh2006,Mosendz2010,Ominato2025}. This versatility has made spin pumping a widely used route for characterizing spin–charge conversion and interfacial spin transport \cite{Mosendz2010,Haertinger2015,Sahu2024}.

Recently, it has been shown that magnetization dynamics can induce strongly nonlinear transport responses, including the generation of ultrahigh harmonics in spin and charge pumping when additional symmetry-breaking mechanisms are present \cite{Ly2022, Ly2025}. Such high-harmonic spin-current emission naturally connects spin pumping to the broader field of ultrafast and nonlinear spintronics, where spin currents driven on femtosecond to picosecond timescales enable efficient terahertz radiation and frequency conversion \cite{Kampfrath2013NatNano,Seifert2016NatPhoton,Leitenstorfer2023THzRoadmap,Kampfrath2023GuestEditorial}. From this perspective, spin pumping emerges as a platform at the intersection of high-harmonic generation (HHG)\cite{Goulielmakis2022} and spintronic terahertz emission \cite{Papaioannou2021}, underscoring the intrinsically multidisciplinary nature of dynamical carrier transport phenomena.

In its conventional realization, spin pumping is driven by ferromagnetic \cite{Tserkovnyak2002} or antiferromagnetic \cite{Cheng2014} resonance, and the emitted spin current oscillates at the same frequency as the driving dynamics. Symmetry considerations prohibit direct charge transport in this setting, such that only pure spin angular momentum is pumped.
{In our previous works, we demonstrated that extremely high harmonics can emerge in carrier pumping in the presence of relativistic mechanisms \cite{Ly2022} or noncollinear magnetic textures \cite{Ly2023}. More recently, the altermagnetic order has been identified as a nonrelativistic platform capable of supporting high-harmonic carrier pumping \cite{Ly2026}, owing to its intrinsically nonrelativistic alternating spin–momentum coupling \cite{Smejkal2020, Satoru2020}.}

In the present work, we propose a distinct route to high-harmonic spin pumping that does not rely on spin-momentum coupling. Building on recent insights into the role of nonlinearities in adiabatic energy spectra \cite{Ly2025}, we show that the essential ingredient underlying high-harmonic pumping is not SOC per se, but rather the presence of intrinsic nonlinear band dynamics. While such nonlinearities often arise in systems with spin–orbit interactions, we demonstrate that they can also be generated in purely magnetic systems. Specifically, we consider a SOC-free magnetic structure in which a second static order parameter is introduced perpendicular to the cone axis of the precessing magnetization. This additional magnetic order qualitatively reshapes the adiabatic energy levels, inducing strong nonlinear features that directly propagate into the dynamical transport response.
As a result, spin pumping accompanied by a cascade of higher harmonics becomes possible even in the complete absence of SOC. Our findings establish a new mechanism for nonlinear and high-frequency spin pumping rooted solely in magnetic structure and dynamics, thereby broadening the class of systems capable of hosting ultrafast spin-transport phenomena beyond the existing spin–orbit-based approaches.

{
\section{From Band Dynamics to Nonlinear Spin Transport}

In conventional spin pumping, a precessing magnetic order placed in proximity to a normal metal injects a spin current into the latter. From a band-structure perspective, this process is governed by the adiabatic evolution of electronic states in a time-dependent magnetic background. To identify the conditions under which this evolution can generate high-harmonic responses in transport, it is sufficient to consider a minimal model consisting of a one-dimensional chain hosting a driven magnetic order. Despite its simplicity, this model provides a transparent platform to connect nonlinear band dynamics to high-frequency spin transport.

The corresponding time-dependent Bloch Hamiltonian reads
\begin{equation}
\label{trivial}
\mathcal{H}(t) = -2\gamma \cos k + J_0 \boldsymbol{m}(t)\cdot\boldsymbol{\sigma},
\end{equation}
where $\gamma$ denotes the nearest-neighbor hopping amplitude, $k$ is the crystal momentum, and $J_0$ is the ferromagnetic $s$--$d$ exchange coupling. The unit vector
\(
\boldsymbol{m}(t) = (\sin\theta\cos\omega t, \sin\theta\sin\omega t, \cos\theta)
\)
describes a uniformly precessing magnetic order with cone angle $\theta$ and angular frequency $\omega$.

This Hamiltonian yields two energy bands,
\begin{equation}
\label{dispersion0}
\varepsilon_{\pm} = -2\gamma\cos k \pm J_0,
\end{equation}
which are strictly time independent. As a consequence, the band structure itself does not encode any nonlinear temporal dynamics, and HHG is absent at the spectral level. Although the eigenstates acquire a harmonic time dependence through the rotating magnetization, this only results in a homodyne response in the pumped spin current, confined to the fundamental driving frequency.

This observation highlights a key point: the emergence of high harmonics in transport requires the energy bands themselves to acquire a nonlinear time dependence. In previous work \cite{Ly2025}, this condition was realized by introducing Rashba SOC, which reshapes the dispersion in Eq.~\eqref{dispersion0} into a square-root form and leads to strongly nonlinear carrier dynamics. Here, we demonstrate that a similar nonlinear band dynamics can be achieved without invoking SOC, relying instead solely on magnetic order.

To this end, we introduce a secondary static magnetic interaction that is noncollinear with the dynamical precession. If the two magnetic orders were collinear, the time dependence could be eliminated by a gauge transformation to the rotating frame, precluding any nonlinear response. We therefore consider an additional static magnetic order transverse to the precession axis of the driving magnetization. The Hamiltonian in Eq.~\eqref{trivial} then generalizes to
\begin{equation}
\label{nontrivial}
\mathcal{H}(t) = -2\gamma \cos k + J_0 \boldsymbol{m}(t)\cdot\boldsymbol{\sigma} + J_1 \sigma_x,
\end{equation}
where $J_1$ denotes the strength of the static transverse ferromagnetic moment.

The resulting energy dispersion reads
\begin{equation}
\varepsilon_{\pm} = -2\gamma\cos k \pm \color{blue}\sqrt{D(t)}.
\end{equation}
{Where we have introduced the dynamical term 
\begin{equation}
\label{eq:d}
D(t)=J_0^2+J_1^2+2 J_0 J_1 \sin \theta \cos\omega t
\end{equation}

The energy levels now exhibit an explicit nonlinear dependence on time, through the dynamical term Eq.~\eqref{eq:d}.} This nonlinear band dynamics constitutes the central ingredient of the present mechanism: it directly imprints higher harmonics onto the carrier motion and is therefore expected to generate a highly nonlinear transport response \cite{Ly2025}.

{
\section{Analytical expression for the nonlinear spin current}

To gain further understanding of the emergence of higher harmonics in this noncollinear magnetic configuration, we exploit the adiabatic spin pumping theory \cite{Brouwer1998, Tserkovnyak2002b}. In fact, the adiabatic parametric pumping theory remains valid as long as the frequency is assumed to be very small with respect to the magnetic gap. According to the standard spin pumping theory, the spin current associated with the precession of a unit vector ${\boldsymbol{m}}$, parametrized by the adiabatic phase $\phi=\omega t$ is given as \cite{Tserkovnyak2002b}:
\begin{equation}
\label{Ivector}
\boldsymbol{\rm{I}}^{\sigma}(t) = \frac{\hbar}{4\pi}\big(A_r\, \boldsymbol{m}(t) \times \dot{\boldsymbol{m}}(t) - A_i\, \dot{\boldsymbol{m}}(t)\big),
\end{equation}
where $A_r$ and $A_i$ are real and imaginary parts of the spin-pumping conductance \cite{Tserkovnyak2002b}, which is related to spin-mixing conductance \cite{Brataas2000}. 

In the presence of the transversal order the magnetic moment becomes

\begin{equation}
{\boldsymbol{m}}_{\Sigma}(t) = J_0 \boldsymbol{m}(t) + J_1 \hat{\boldsymbol{x}},
\end{equation}
Yet, to apply the standard pumping expression Eq. (\ref{Ivector}), one should construct the corresponding magnetic unit vector as

\begin{equation}
\tilde{\boldsymbol{m}}(t) = {\boldsymbol{m}}_{\Sigma} / |{\boldsymbol{m}}_{\Sigma}|,
\end{equation}
One finds
\begin{equation}
\tilde{\boldsymbol{m}}(t) = \boldsymbol{m}/\sqrt{D}.
\end{equation}

{In this case, the amplitude of the effective magnetic moment becomes time-dependent. However, one can always define time-dependent spin projection matrices $$\hat{\boldsymbol{u}}^{\uparrow,\downarrow}=\frac{1}{2}(\hat{\boldsymbol{1}}\pm \boldsymbol{\sigma}\cdot \tilde{\boldsymbol{m}}(t)),$$  
similarly to the time-independent amplitude case. 
This justifies the applicability of the standard spin pumping theory \cite{Tserkovnyak2002b} to the modified time-dependent order $\tilde{\boldsymbol m}(t)$. As a consequence, the resulting adiabatic scattering matrix would be of highly non-linear nature, due to the $\sqrt{D}$ dependence in $\tilde{\boldsymbol{m}}$.}

The corresponding modified spin-current vector is obtained as

\begin{equation}
\label{eq:itilde}
\begin{split}
\frac{4\pi}{\hbar}\tilde{\boldsymbol{\rm{I}}}^{\sigma}(t) 
&= A_r\frac{ J_0^2}{D}\boldsymbol{m} \times \dot{\boldsymbol{m}}  - A_i\frac{J_0}{\sqrt{D}}\dot{\boldsymbol{m}}
             \\
            &+A_r \frac{J_0 J_1}{D}{\hat{\boldsymbol{x}}\times\dot{\boldsymbol{m}}}+ {A_i}\frac{\omega J_0 J_1\sin\theta\sin\omega t}{D^{3/2}}~\boldsymbol{m}_{\Sigma} 
\end{split}
\end{equation}
}
{This shows that the resulting spin current bears higher order harmonics encoded in the nonlinear terms proportional to $D^{-1}$, $\sqrt{D}^{-1}$ and $\sqrt{D}^{-3}$. 
We shall emphasize the emergence of the term proportional to $\boldsymbol{m}$, which is absent in the standard pumping $J_1=0$. This shows explicitly how the transversal correction modifies the standard spin current to enable for the emergence of HHG in the underlying spin current. 
}

{ It is important to note that this analytical description cannot be utilized in a quantitative manner for the more realistic geometry where the dynamical order and static field reside in different regions of the sample (as depicted in Fig.~\ref{fig:sketch}).
Nevertheless, it demonstrates transparently the emergence of the nonlinearities as these two noncollinear orders interact.
}

{In the following, we perform numerical simulations of spin transport in the corresponding magnetic heterostructure to complement the analytical description provided above.}

{
\begin{figure}
    \includegraphics[width=0.5\textwidth]{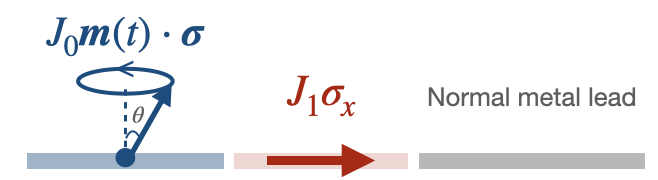}
    \caption{Schematic of the simulated geometry, comprising a precessing ferromagnetic order (blue cone) adjacent to a static magnetic order (red arrow) oriented perpendicular to the precession axis of the dynamical magnetization. The system is connected to a normal-metal lead (gray region), into which spin currents are pumped.}
    \label{fig:sketch}
\end{figure}

\section{Numerical Results}

\begin{figure}
    \includegraphics[width=0.5\textwidth]{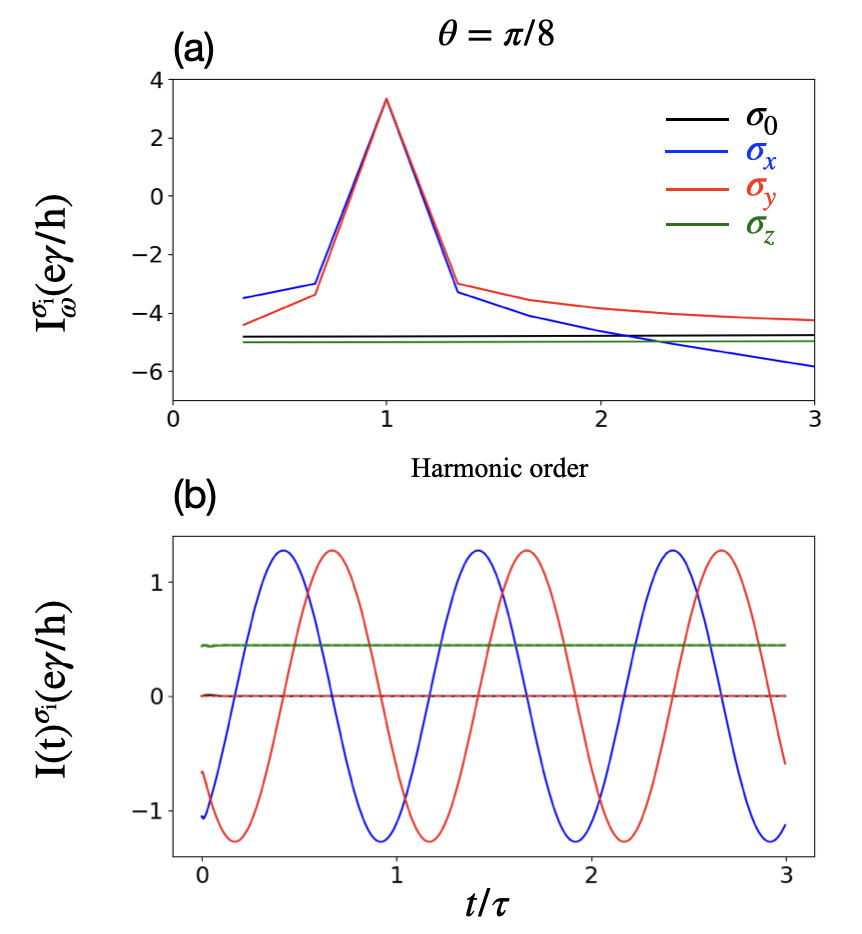}
    \caption{Charge and spin currents for standard spin pumping ($J_0=2\gamma$, $J_1=0$). Panel (a) shows the Fourier spectra of the spin ($\rm{I_{\omega}^{\sigma_x, \sigma_y, \sigma_z}}$) and charge ($\rm{I_{\omega}^{\sigma_0}}$) currents, while panel (b) displays the corresponding time-domain signals. The calculation is performed at Fermi energy $E_F=-1.8\gamma$ with driving frequency $\omega=0.01\,\gamma/\hbar$ and precession cone angle $\theta=22.5^{\circ}$.}
    \label{fig1}
\end{figure}

To complement the analysis of the energy spectrum presented above, we now turn to numerical simulations of the time-dependent transport response generated by the Hamiltonian in Eq.~\eqref{nontrivial}. In particular, we consider a one-dimensional chain hosting a driven magnetic order in addition to a perpendicular static magnetic order (see Fig. \ref{fig:sketch}). We compute the resulting spin and charge currents flowing into an adjacent normal-metal lead (depicted in gray in Fig. \ref{fig:sketch}). This approach allows us to directly access the full time-resolved transport response beyond analytical limits and to validate the emergence of nonlinear pumping signatures discussed previously.

\begin{figure*}
    \includegraphics[width=\textwidth]{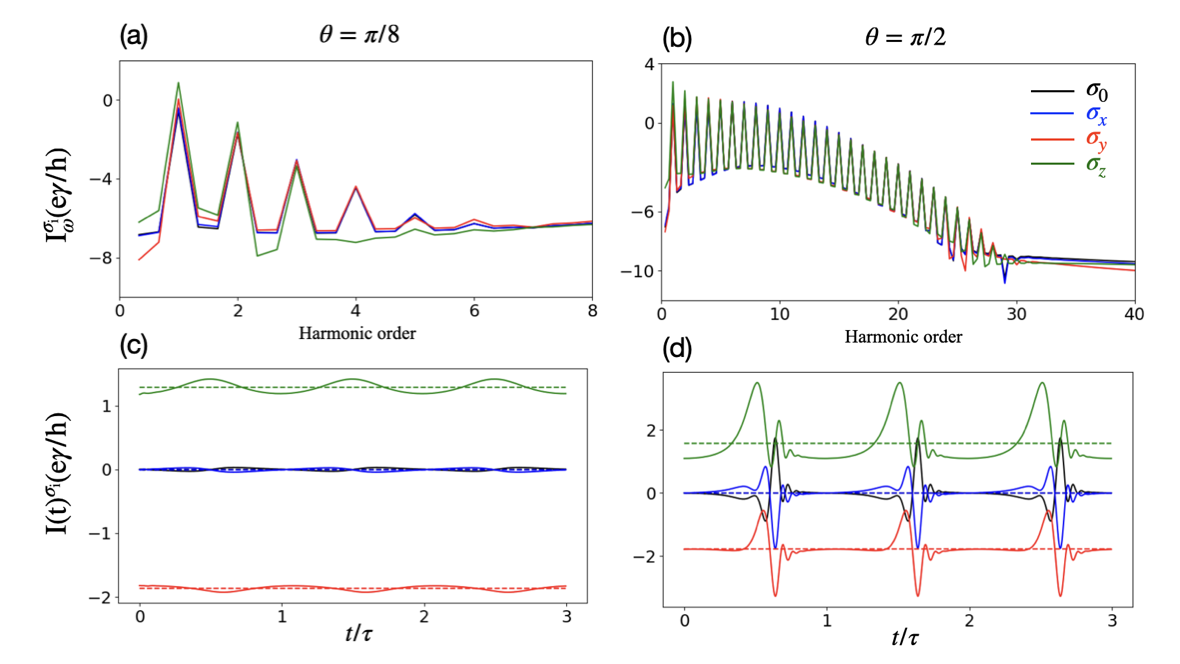}
    \caption{Charge and spin currents in the presence of a static transverse magnetic order. Panels (a) and (c) show the Fourier spectra and time-domain signals, respectively, for a precession cone angle $\theta=\pi/8$. Panels (b) and (d) display the same quantities for the in-plane precession configuration $\theta=\pi/2$. The exchange couplings are {$J_0=2\gamma$ and $J_1=1.5\gamma$}. The Fermi energy and driving frequency are the same as in Fig.~\ref{fig1}.}
    \label{fig2}
\end{figure*}

The transport calculations are performed using the non-equilibrium quantum transport platform \cite{tkwant}, which enables the simulation of time-dependent scattering problems. The scattering wave function is first obtained at time $t=0$ using the stationary transport solver \cite{kwant}, and subsequently evolved according to the time-dependent Schr\"odinger equation. The spin current polarized along direction $j$, flowing at time $t$ from a site $i$ in the chain to a neighboring site $l$ at the lead interface, is evaluated as
\begin{equation}
{\rm{I}}^{\sigma_j}(t) = {2}\Im\!\left[\psi_i^\dagger(t)\, \mathcal{H}_{il}(t)\, \sigma_j\, \psi_l(t)\right],
\end{equation}
where $\mathcal{H}_{il}(t)$ denotes the Hamiltonian matrix element between sites $i$ and $l$, and $\psi_i(t)$ is the scattering wave function at site $i$ and time $t$. The charge current is obtained by replacing $\sigma_j$ with the identity matrix.

We begin by considering the standard spin-pumping regime, corresponding to the absence of the static transverse magnetic field ($J_1=0$). Figure~\ref{fig1}(b) displays the time-domain signals of the spin and charge currents in this limit. As expected, the charge current vanishes identically, while the transverse spin components oscillate harmonically at the driving frequency and exhibit no DC component. In contrast, only the spin current polarized along the $z$ direction acquires a finite time-averaged value. These results are in full agreement with the adiabatic theory of spin pumping, {discussed above (see Eq.(\ref{Ivector})).} 

We now turn to the case where the static transverse magnetic field is included. Figure~\ref{fig2} shows the time-domain signals and corresponding Fourier spectra for two different precession cone angles. In agreement with spin-pumping theory, the DC charge current remains zero [see Figs.~\ref{fig2}(b) and \ref{fig2}(d)]. However, the presence of the transverse field gives rise to a strongly nonlinear dynamical response, as evidenced by the emergence of higher harmonics in the Fourier spectra shown in Figs.~\ref{fig2}(a) and \ref{fig2}(c).

Note that the DC spin current polarized parallel to the static field vanishes. In the absence of SOC, spin dynamics in the static-field region is governed by coherent precession about the field axis, which does not mix transverse spin components into the longitudinal one. As the pumped spin current injected from the dynamical region carries no DC polarization along the static-field direction, the corresponding spin component remains uncoupled and decays diffusively. Consequently, in the periodic steady state, the DC spin current parallel to the static field is strictly zero.
In contrast, the spin currents polarized along the $y$ and $z$ directions acquire finite DC contributions. For a small precession cone angle, $\theta=\pi/8$, only a limited number of harmonics is observed. When the dynamical magnetization maximally couples to the secondary magnetic order, corresponding to $\theta=\pi/2$, the harmonic spectrum extends up to the 30th order [Fig.~\ref{fig2}(c)]. 
{ It can be observed that the resulting DC amplitudes depicted as horizontal dashed lines in Fig.~\ref{fig2}(c), are larger than the DC components in the standard pumping case (shown in Fig.~\ref{fig1}(b)), but these amplitudes remain in the same order of magnitude. This suggests that the effect can be observed using similar techniques employed for standard spin-pumping measurements.}

Beyond the emergence of higher harmonics, an additional hallmark of the nonlinear dynamics is encoded in the frequency dependence of the DC transport response. In particular, the DC components of the pumped spin current are expected to exhibit a nonlinear scaling with the driving frequency $\omega$. Figure~\ref{fig3} shows the amplitudes of the DC spin currents as functions of $\omega$. In the standard spin-pumping limit ($J_1=0$), the DC spin current is a linear combination of $\boldsymbol{m}(t)\times\dot{\boldsymbol{m}}(t)$ and ${\boldsymbol{m}}(t)$ (Eq. \ref{Ivector}). Therefore, it scales linearly with the frequency $\omega$. In contrast, once the transverse magnetic order is introduced, this linear scaling no longer holds. As seen for both spin-current components polarized along the $y$ and $z$ directions in Fig.~\ref{fig3}, the dependence of the DC response on $\omega$ becomes increasingly nonlinear as the precession cone angle is increased. This nontrivial frequency dependence provides a clear and experimentally accessible signature of the nonlinear spin pumping induced by a noncollinear configuration of precessing and static magnetic orders.

Furthermore, a similar nonlinear dependence on the precession cone angle $\theta$ is expected and constitutes an additional hallmark of the nonlinear spin dynamics. Our simulations indicate that the high-harmonic response is strongly enhanced for large precession cone angles. While $\theta$ is typically small in conventional spin-pumping experiments, an alternative and particularly promising platform has recently emerged. Indeed, it has been shown that when magnetization dynamics are driven by microwave oscillating voltages, a large-angle precession can be sustained by appropriately tuning the microwave frequency \cite{Yamamoto2020,Imamura2020}. Such a platform offers an attractive route for optimizing high-harmonic spin pumping in magnetic systems and provides favorable conditions for the experimental observation of purely magnetic, high-harmonic spin pumping.
}

\begin{figure*}
    \includegraphics[width=\textwidth]{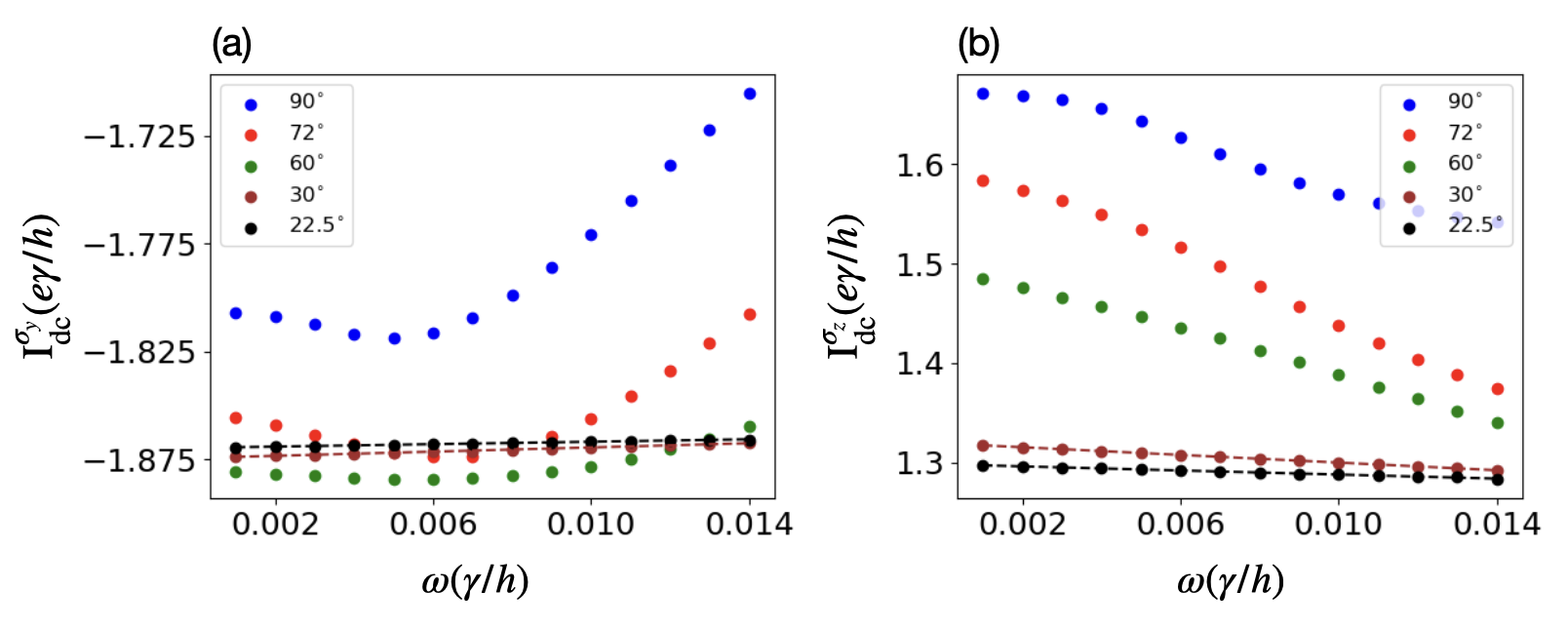}
    \caption{Calculated DC components of the pumped spin current polarized along the $y$ [(a)] and $z$ [(b)] directions are shown as functions of the driving frequency $\omega$ for precession cone angles ranging from $\theta=\pi/8$ to $\pi/2$. The pumped currents are evaluated for the same set of parameters as in Fig.~\ref{fig2}. The dashed brown and black lines correspond to linear fits of the underlying data.
  }
    \label{fig3}
\end{figure*}

\section{Discussion and Conclusion}
The numerical simulations presented above demonstrate that introducing a static magnetic order transverse to a driven magnetization qualitatively alters the nature of spin pumping. While standard spin pumping is characterized by a purely harmonic response with vanishing charge current and a single-frequency spin signal, the additional transverse magnetic field renders the adiabatic energy spectrum explicitly time dependent and strongly nonlinear. This nonlinearity directly manifests itself in the transport response through the emergence of higher harmonics in the spin currents.

Importantly, the appearance of high harmonics does not rely on SOC, but instead originates from the interplay between two noncollinear magnetic orders. The resulting nonlinear time dependence of the instantaneous energy levels constitutes the minimal ingredient required for HHG generation in the pumping response. {While a minimal one-dimensional model has been considered in this work, the emergence of HHG within a 
similar noncollinear setup in a two dimensional magnetic structure is presented in the Supplemental Materials.}
As evidenced by the numerical results, the strength and extent of the harmonic spectrum are controlled by the relative orientation of the two magnetic orders, with maximal harmonic generation occurring when the dynamical magnetization most strongly couples to the static transverse field.

Another notable feature of this mechanism is the emergence of DC spin components in selected polarization channels, while the DC charge current vanishes. This behavior clearly distinguishes the present nonlinear magnetic pumping regime from conventional adiabatic spin pumping and provides an experimentally accessible signature of the underlying nonlinear dynamics.

{
Regarding experimental realization, a conventional approach to measuring spin pumping is via the inverse spin Hall effect (ISHE) \cite{Saitoh2006,Mosendz2010}, where the pumped spin current is converted into a DC voltage by contacting the magnetic system with a heavy-metal layer, such as platinum. 
{This can be realized in garnet-platinum heterostructures or iron-nickel compounds combined with other ferromagnets \cite{Ando2011, Swain2024}. Interestingly, a non-collinear configuration can also be achieved in systems incorporating antiferromagnetic order. Since antiferromagnets operate at a different frequency range (typically in the THz regime), one magnetic order can be excited without altering the other. This configuration can be implemented in conventional ferro-antiferromagnetic heterostructures  \cite{Radu2008}, combined with a heavy metal. Another promising platform for observing the effect is the recently investigated magnetic heterostructures for superconducting spin pumping, where a drastic enhancement of spin pumping has been reported \cite{Jeon2018, Jeon2019}.}

However, observing the time-domain response remains challenging and may require further experimental refinement. Interestingly, the emergence of these harmonics within the energy levels dynamics suggests they could be detected using angle-resolved photoemission spectroscopy (ARPES) \cite{Zhong2022}.
}

{We emphasize that the present mechanism leads to the enhancement of the dynamics frequency by roughly an order of magnitude for moderate precession angles. 
Given that conventional ferromagnets typically operate at tens of GHz, our results indicate that the high harmonic spectrum can span several hundred GHz.
}

Taken together, our results establish a conceptually distinct route to high-harmonic spin pumping driven purely by magnetic interactions. Although demonstrated here for uniform magnetic dynamics, the mechanism naturally extends to spatially extended magnetic textures in $d$ dimensions, where the interplay between temporal dynamics and spatial variations of the magnetic order introduces additional nonlinearities. These nonlinearities enable the redistribution of spectral weight from the fundamental drive frequency to higher harmonics, thereby enriching the spin-pumping response. More generally, the coexistence of multiple magnetic regions and time-dependent exchange fields suggests that nonlinear spin transport of this kind should be realizable in a broad class of magnetic systems, providing a promising platform for ultrafast spintronic functionalities and high-frequency signal generation.

\bibliography{refs}
\end{document}